# Harmonic generation from gold nanolayers: an old problem under a new light

L. Rodríguez-Suné,[1,*] J. Trull,[1] C. Cojocaru,[1] N. Akozbek,[2] D. de Ceglia,[3] M. A. Vincenti,[4] M. Scalora[5,**]

[1]*Department of Physics, Universitat Politècnica de Catalunya, Rambla Sant Nebridi 22, 08222 Terrassa, Spain*
[2]*AEgis Technologies Inc., 401 Jan Davis Dr., Huntsville, Alabama 35806, USA*
[3]*Department of Information Engineering – University of Padova, Via Gradenigo 6/a, 35131 Padova, Italy*
[4]*Department of Information Engineering – University of Brescia, Via Branze 38, 25123 Brescia, Italy*
[5]*Aviation and Missile Center, US Army CCDC, Redstone Arsenal, AL 35898-5000 USA*
[*]laura.rodriguez.sune@upc.edu, [**]michael.scalora.civ@mail.mil

**Abstract:** Understanding how light interacts at the nanoscale with metals, semiconductors, or ordinary dielectrics is pivotal if one is to properly engineer nano-antennas, filters and, more generally, devices that aim to harness the effects of new physical phenomena that manifest themselves at the nanoscale. We presently report experimental results on second and third harmonic generation from 20nm- and 70nm-thick gold layers, for TE- and TM-polarized incident light pulses. We highlight and discuss for the first time the relative roles bound electrons and an intensity dependent free electron density (hot electrons) play in third harmonic generation. While planar structures are generally the simplest to fabricate, metal layers that are only a few nanometers thick and partially transparent are almost never studied. Yet, transmission offers an additional reference point for comparison, which through relatively simple experimental measurements affords the opportunity to test the accuracy of available theoretical models. Our experimental results are explained well within the context of the microscopic hydrodynamic model that we employ to simulate second and third harmonic conversion efficiencies, and to simultaneously and uniquely predict the nonlinear dispersive properties of a gold nanolayer under pulsed illumination. Using our experimental observations and our model, based solely on the measured third harmonic power conversion efficiencies we predict $\chi_\omega^{(3)}(1064\text{nm}) \approx (-3.7 + i\,0.21) \times 10^{-18} \,(\text{m/V})^2$, triggered mostly by hot electrons, without resorting to the implementation of a z-scan set-up.

## 1. Introduction

Currently, nanostructures are routinely fabricated and integrated in different photonic devices for a variety of purposes and applications. At the nanoscale, which we interpret to mean that material features may be only a few atomic diameters in size, light-matter interactions can display new phenomena, and conventional approximations may not always be applicable. This is the case for processes like second harmonic generation (SHG) and third harmonic generation (THG). Traditionally, SHG and THG have been studied under phase-matched conditions, primarily to improve efficiency in bulk, optically thick materials (hundreds of microns or millimeters) with high nonlinearities and low material absorption across a spectral range that includes the pump and its harmonics. Under these circumstances, the leading nonlinear polarization term corresponds to the bulk contribution from the nonlinear potential described by dipolar second and third order nonlinear susceptibility tensors ($\chi^{(2)}$ and $\chi^{(3)}$). However, at the nanoscale and under pulsed illumination, the effective $\chi^{(2)}$ and $\chi^{(3)}$ may not coincide with their bulk, local counterparts, may depend on pulse duration and on the type of nonlinearity that is triggered and their associated linear and nonlinear dispersions. In addition, contributions to the nonlinear polarizations arising from electric quadrupole-like and magnetic sources should also be taken into account.

Interest in the nonlinear optical properties of metals arches back to the beginning of nonlinear optics, is still a subject of debate, and is now more relevant than ever insofar as nano-plasmonics is concerned. From a historical point of view, it was pointed out in reference [1] that in centrosymmetric media, where bulk $\chi^{(2)}$ vanishes, SHG arises from a magnetic dipole term and from an electric quadrupole-like contribution. Later, in reference [2], it was shown that the quadrupole source term was equivalent to a nonlinear surface contribution, and it was proposed that SHG in metals may be explained by considering separate bulk and surface contributions having different weights associated with free electron dynamics. Early experimental evidence of SHG in metals may be found in reference [3], where reflected SHG was measured from a silver (Ag) mirror. In references [4] and [5] experimental results of reflected SHG were reported for Ag and gold (Au) mirrors. Additional theoretical and experimental studies followed, a small sample of which may be found in references [6-8], where the main approach is to separate and distinguish between surface and volume contributions. For example, based on the idea of identifying surface and volume sources, in reference [9] the magnitude and relative phase of second-order susceptibility tensor elements may be determined for thin-film metal samples of silver, gold, copper, aluminum and tantalum. As another example, in reference [10] SHG is reported in diffraction from arrays of symmetric Au nanoparticles, while SHG from single Au elliptical nanoparticles is experimentally investigated in reference [11]. Other experimental studies of SHG from Au are reported in references [12] and [13]. Third order nonlinearities have also been studied in a variety of Au samples in various geometries [14-19].

As mentioned above, most current theoretical models of SHG rely on the introduction of phenomenological or effective surface and bulk parameters that generally lack a detailed, microscopic, dynamical description of light propagation and light-matter interactions. Here, we present experimental measurements of the angular dependence of SHG and THG in transmission and reflection from 20nm- (transparent) and 70nm-thick (opaque) Au nanolayers deposited on a fused silica substrate. These experimental observations are compared with numerical simulations based on a theoretical model that embraces full-scale, time-domain coupling of matter to the macroscopic Maxwell's equations. Our approach consists in formulating a microscopic, hydrodynamic model in order to understand linear and nonlinear optical properties of metals by accounting for competing surface, magnetic, and bulk nonlinearities arising from both free and bound (or valence) electrons. Just as importantly, our model preserves linear and nonlinear material dispersions, nonlocal effects, and the influence of hot electrons, i.e. electrons that may be temporarily excited from the valence into the conduction band, thus causing a transient increase of the free electron density. This model is adapted and applied anew based on previous work reported in references [20-23], where it is described in details. Application of Newtonian dynamics to conduction and valence electrons leads to the following simultaneous material equations of motion in three spatial coordinates and time:

$$\ddot{\mathbf{P}}_f + \tilde{\gamma}_f \dot{\mathbf{P}}_f = \frac{e^2 \lambda_0^2 n_{0,f}}{m_0^* c^2}\mathbf{E} + \tilde{\Lambda}(\mathbf{E}\bullet\mathbf{E})\mathbf{E} - \frac{e\,\lambda_0}{m_0^* c^2}(\nabla\bullet\mathbf{P}_f)\mathbf{E} + \frac{e\,\lambda_0}{m_0^* c^2}\dot{\mathbf{P}}_f \times \mathbf{H}$$
$$+ \frac{3E_F}{5m_0^* c^2}\left(\nabla(\nabla\bullet\mathbf{P}_f) + \frac{1}{2}\nabla^2 \mathbf{P}_f\right) - \frac{1}{n_{0,f}e\lambda_0}\left[(\nabla\bullet\dot{\mathbf{P}}_f)\dot{\mathbf{P}}_f + (\dot{\mathbf{P}}_f\bullet\nabla)\dot{\mathbf{P}}_f\right] \quad (1)$$

$$\ddot{\mathbf{P}}_{bj} + \tilde{\gamma}_{bj}\dot{\mathbf{P}}_{bj} + \tilde{\omega}_{0,bj}^2 \mathbf{P}_{bj} - \tilde{\beta}(\mathbf{P}_{bj}\bullet\mathbf{P}_{bj})\mathbf{P}_{bj} = \frac{n_{0,b}e^2\lambda_0^2}{m_{bj}^* c^2}\mathbf{E} + \frac{e\,\lambda_0}{m_{bj}^* c^2}(\mathbf{P}_{bj}\bullet\nabla)\mathbf{E} + \frac{e\,\lambda_0}{m_{bj}^* c^2}\dot{\mathbf{P}}_{bj}\times\mathbf{H}\cdot \quad (2)$$

$\mathbf{P}_f$ and $\mathbf{P}_{bj}$ are free and bound electron polarizations, respectively. These equations are solved simultaneously together with the macroscopic Maxwell's equations, where the total polarization is the vectorial sum of all polarization components. The spatial and temporal derivatives in Eqs.(1-2) are carried out with respect to the following scaled coordinates: $\xi = z/\lambda_0$, $\varsigma = y/\lambda_0$, and $\tau = ct/\lambda_0$, where $\lambda_0$ is a reference wavelenght. The fields are assumed to be invariant along the transverse *x*-direction. SHG arises mostly via Eq.(1), although Eqs.(2) (the *j* counter in Eqs.(2) indicates multiple bound electron species) are also a source of SHG. THG on the other hand is triggered by both Eqs.(1) and (2) via the two discernible third order terms. Free electrons are under the influence of electric, magnetic, and convective forces, as well as nonlocal effects. The instantaneous free electron density is allowed to vary as a function of applied intensity, changes that are exemplified by the term $\tilde{\Lambda}(\mathbf{E} \bullet \mathbf{E})\mathbf{E}$. $\tilde{\Lambda}$ is a scaled coefficient that generally depends on the equilibrium free electron plasma frequency, conductivity (which is also a function of wavelength), temperature, and pulse duration [23]. In a metal like Au, one free (Drude) and two bound (Lorentz) electron species generally suffice to describe the local, linear dielectric permittivity down to a wavelength of approximately 200nm. For simplicity, each Lorentz species is characterized by a third order, isotropic nonlinearity $\mathbf{P}_{bj} = -\tilde{\beta}(\mathbf{P}_{bj} \bullet \mathbf{P}_{bj})\mathbf{P}_{bj}$, where the scaled coefficient $\tilde{\beta} \approx \left(\omega_{0,b1}^2 + \omega_{0,b2}^2\right)\lambda_0^2 / \left(2L^2 n_{0b}^2 e^2 c^2\right)$ may be derived from a nonlinear, two-dimensional oscillator model. $\omega_{0,b1}$ and $\omega_{0,b2}$ are the two resonance frequencies, *L* the lattice constant, $n_{0b}$ the bound electron density, $\lambda_0 = 1\mu m$, and *c* is the speed of light in vacuum. In our case, $n_{0b} \sim 5.8 \times 10^{22} / cm^3$ (similar to the free electron density, although lower d-shell electrons may also contribute at short-enough wavelengths [24]), the lattice constant $L \sim 3 \times 10^{-8} cm$, while $\omega_{0,bj} \approx 10^{16}/\sec$ combine to yield an average resonance wavelength between 300-400nm. As a result, atomic size, crystal structure, and linear oscillator parameters effectively predetermine the dimensionless coefficient $\tilde{\beta} \approx 10^{-8}$, which in turn governs *all* third order effects triggered by the background crystal, i.e. bound electrons, including self-phase modulation, nonlinear absorption, and THG power conversion efficiencies. At the same time, electrons are excited into the conduction band at a rate that depends on incident power density via the strength of the coefficient $\tilde{\Lambda}$, which for our purposes may be assumed to be approximately constant. This third order effect adds considerably to, and can even dominate THG. Our measured efficiencies in fact suggest that the TH signal originates mostly from hot electrons. In turn, applying our model leads us to conclude that $|\chi_\omega^{(3)}| \approx 10^{-18} - 10^{-17} (m/V)^2$. We describe the results below. To our knowledge, the model exemplified by Eqs.(1-2) has never before been used in this context, and is unique in several additional respects. For more details on the basic foundations of the model we refer the reader to references [20-23].

## 2. Samples and experimental set-up

The Au films were deposited from an Au target using a magnetron sputtering system. The DC target power was 100W and deposition pressure 3mTorr. The films were analyzed with both a spectrophotometer and spectroscopic ellipsometer for thickness calibration and optical constants. We performed the experiments in a set-up capable of measuring both transmitted and reflected SH and TH signals generated by the pump as it traverses the Au nanolayers. A schematic representation of this set-up is shown in Fig. 1. Two different sources were used in our experiments. The first one was a Ti:Sapphire oscillator at 800nm (Coherent - MIRA), emitting 140fs pulses at 76MHz repetition rate, with a CW output average power of 1.7 W, corresponding to approximately 22 nJ/pulse. The second source was a pulsed fiber laser (FYLA PS50) emitting a train of 13ps pulses at 1064nm, with a CW average output power of 2W at 1MHz repetition rate, delivering pulses of 2μJ/pulse. The polarization (TE or TM) of the fundamental beam (FF) incident on the sample was controlled using a half-wave plate. Any possible SH and TH signals arising from different optical components (lenses, mirrors, etc.) placed along the set-up before the sample were removed by spectral filters. To obtain the desired fundamental beam intensities on the sample in the range of 1-2GW/cm$^2$, we focused the fundamental beam at the sample plane by means of plano-convex lenses with focal length f=200mm for the SH measurements, and f=100mm for the TH experiments. The sample was mounted on a rotary support arm that ensures precise control of the FF incident angle, allowing measurements of the SH or TH signals as a function of the fundamental beam angle of incidence. The most critical part of the setup is the detection arm, since the expected power conversion efficiencies are in the range $10^{-8}$-$10^{-11}$. Immediately after the sample, a filter attenuating the FF radiation was used, thus avoiding any potential SH or TH generation from the surfaces of the optical elements in this part of the set-up. An anti-reflection coated lens with focal length f=100mm was used to collimate the beam after the sample, and a polarizer served to analyze the polarization state of the SH or TH generated fields. In order to detect the faint SH or TH signals arising from the Au samples, it was crucial to separate them from the residual fundamental beam. To this end, we used a prism and a blocking edge to separate and obscure the remaining fundamental field radiation from the SH or TH path. Finally, a photomultiplier tube is used to detect the harmonic signals, on which we place a narrow-band spectral filter having a 20nm band pass transmission around either the SH or the TH wavelength.

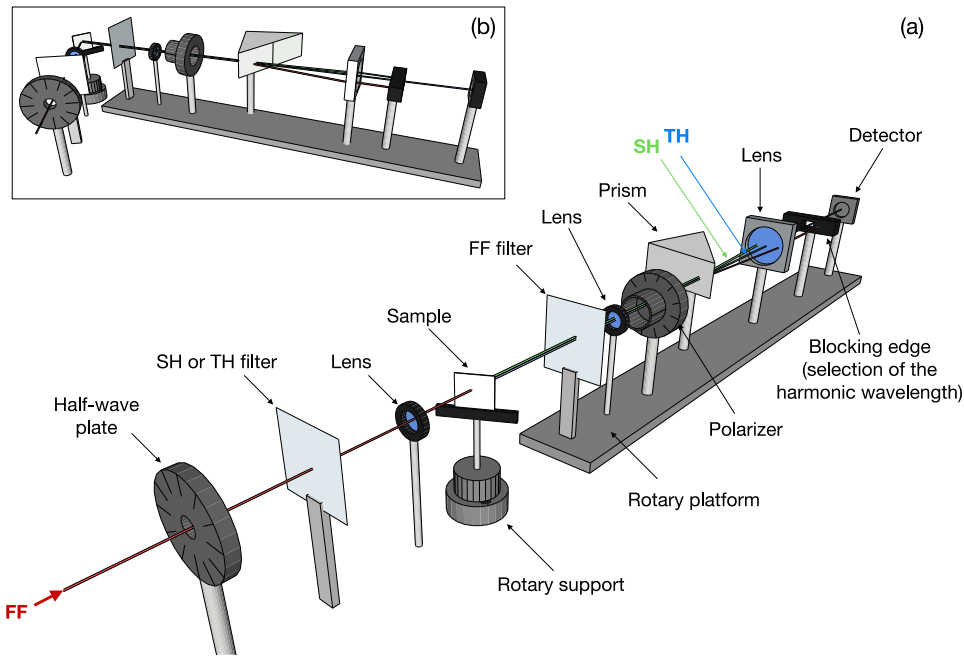

Fig. 1. Set-up developed to measure SH and TH signals generated when light tuned at 800nm or 1064nm interacts with an Au nanolayer. The detection system is mounted on a rotary platform so measurements can be taken in transmission (a) and in reflection (b).

For a more accurate detection, the signal reaching the photomultiplier is modulated in different ways, depending on the laser source. For the femtosecond pulses at 800nm and at 76MHz repetition rate, we use a mechanical chopper modulating the fundamental signal at typical frequencies of the order of 100Hz with a duty cycle of 50%. The picosecond source at 1064nm can be modulated so it delivers a train of N pulses separated by 1 microsecond (1MHz) which are repeated at a rate of 1kHz. In this way we can integrate on the photomultiplier signal measurement the response of N pulses and obtain a larger signal. The entire detection system is mounted on a rotary platform that allows us to precisely set the detection angle and to take measurements in transmission and in reflection, as shown in Fig.1(a) and Fig.1(b), respectively. Our set-up is able to detect SH and TH signals having power conversion efficiencies of order $10^{-8}$ down to $10^{-14}$. We carried out a detailed calibration procedure to accurately estimate the efficiency of a given process as the ratio between the SH or TH intensity generated in transmission or reflection, and the total peak pump pulse intensity just before the sample. The calibration procedure also included the measurement of possible harmonic signals coming from the blank substrate of fused silica.

### 3. Experimental and theoretical results

Transmitted and reflected SH efficiencies are measured at a fixed wavelength as functions of fundamental incident angle for two different samples of 20nm- and 70nm-thick Au layers. The generated SH signals at 532nm and 400nm are recorded for incident pulses having carrier wavelengths tuned to 1064nm (pulse duration ~13ps) and 800nm (pulse duration ~140fs), respectively. In either case the simulations are carried out using 100fs pulses. Barring the intervention of additional physical phenomena, like slow nonlinear thermal effects, the spectral response converges rapidly as a function of incident pulse duration. THG was measured by illuminating the 20nm-thick Au layer with the pulsed 1064nm source, which generates a TH field at 355nm.

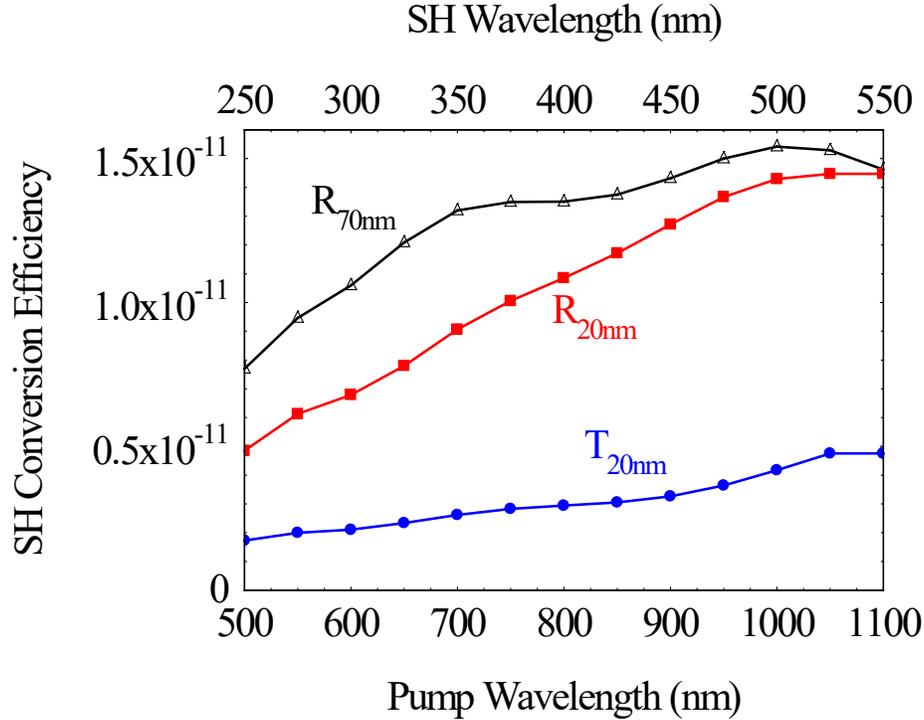

Fig. 2. Calculated SHG conversion efficiency vs. incident wavelength for 20nm- and 70nm-thick gold layers. Transmitted SHG efficiency for the 70nm-thick layer is negligible and is not shown. 100fs pulses are incident at 55°, and have peak power density of approximately 1GW/cm². We assume free and bound electron masses and densities are $m_{free}=m_{bound}=m_e$, and $n_b\sim5.8\times10^{22}/cm^3$, respectively.

In Fig. 2 we show the simulated spectral SH response in reflection and transmission as a function of incident carrier wavelength for both 20nm- and 70nm-thick gold layers, at a fixed 55° angle of incidence. Transmitted SHG for the thicker layer is negligible and is omitted. For each simulation point shown in Fig. 2, the incident pump field consists of a pulse approximately 100fs in duration. At longer wavelengths both metal layers act increasingly as mirrors, as the real part of the dielectric constant becomes progressively more negative, a fact that leads to an enhanced, nonlinear surface response. The relative importance of surface and volume sources changes as a result of varying degrees of penetration depth, and peak SH response is expected to occur at different angles of incidence.

In Fig. 3 we plot and compare experimentally measured and predicted transmitted SHG efficiencies at 800nm and 1064nm fundamental wavelengths, respectively, as functions of the angle of incidence for the 20nm-thick Au layer. The agreement between predicted and measured values is good and occurs notwithstanding the fact that simulations are carried out using incident 100fs pulses in both instances. This confirms that rapid convergence is achieved as a function of incident pulse duration for flat structures that display no geometrical spectral features. It is evident that experimental and theoretical results agree well in both cases, in shape, amplitude, and peak locations. At longer wavelengths, the main peak shifts to larger angles, field penetration depth decreases and surface charge discontinuities, exemplified by Coulomb ( $-\frac{e\lambda_0}{m_0^* c^2}(\nabla \cdot \mathbf{P}_f)\mathbf{E}$ ) and convective ( $-\frac{1}{n_{0,f} e \lambda_0}\left[(\nabla \cdot \dot{\mathbf{P}}_f)\dot{\mathbf{P}}_f + (\dot{\mathbf{P}}_f \cdot \nabla)\dot{\mathbf{P}}_f\right]$ ) terms, increase and become the main source of SHG, although the magnetic contribution encompasses both nonlinear surface and volume currents.

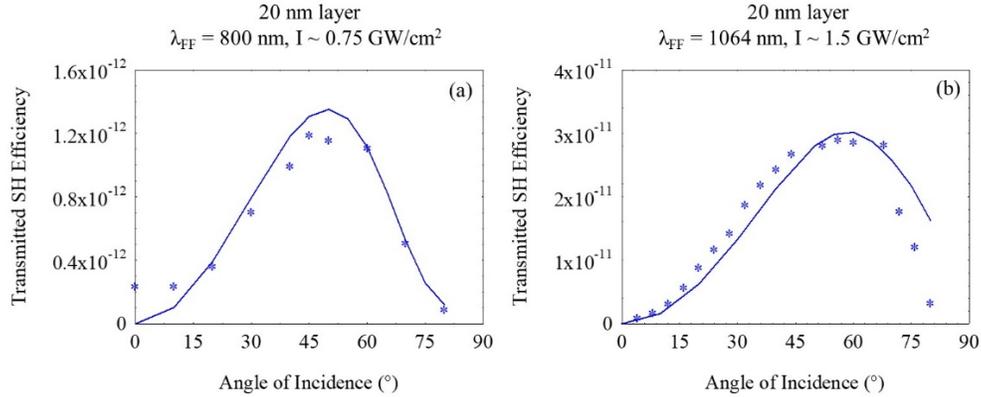

Fig. 3. Measured (blue * markers) and simulated (solid blue curves) of transmitted SHG efficiency as a function of incident angle for the 20nm-thick Au layer when the incident field is tuned at (a) 800nm, with peak power density of approximately 0.75GW/cm$^2$, and (b) 1064nm and peak power density of order 1.5GW/cm$^2$.

In Fig. 4 we show measured and predicted angular dependence of transmitted and reflected SHG conversion efficiencies from the 20nm-thick gold sample for 800nm incident wavelength. Measurements and simulations agree remarkably well, including the ratio of maximum reflected and transmitted efficiencies, and the relative angular positions of the maxima. We emphasize that we have not assumed any effective surface or volume nonlinearities to explain our experimental results. Instead, we rely completely on knowledge of approximate free electron mass and density, which may be ascertained by various means, and incident peak power density. These are the only free parameters in the calculations, which lead to unprecedented, objective agreement between theory and experimental results of SHG from metal structures.

In Fig. 5 we plot reflected SHG power conversion efficiencies as functions of incident angle for 800nm and 1064nm pulsed laser sources for the 70nm-thick gold layer. Transmission is negligible in both cases. We remark that once again experimental and theoretical results agree quite well. Peak SHG performance shifts from ~60° at 800nm, to ~75° at 1064nm, reflecting changes in field penetration depth in the two different spectral ranges.

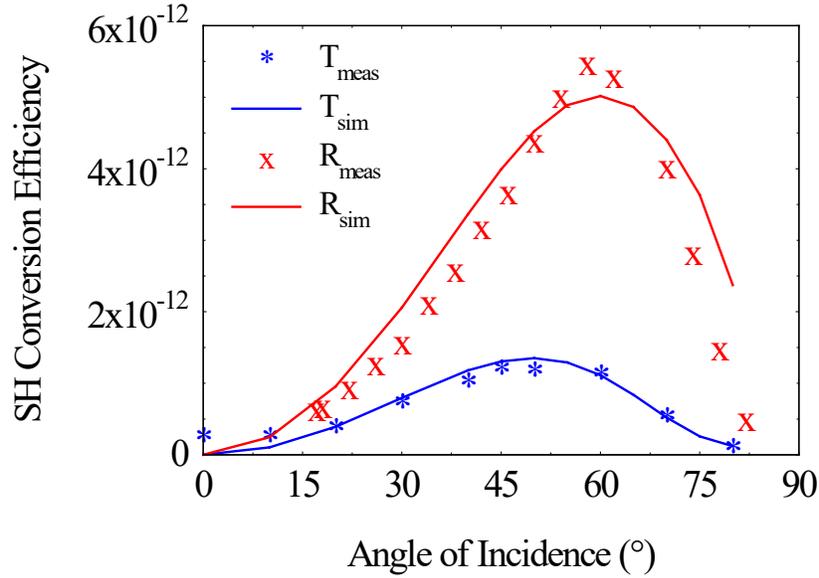

Fig. 4. Measured (blue ∗ and red x symbols) and corresponding simulated (solid blue and red curves) of transmitted and reflected SHG efficiencies, as indicated, as functions of incident angle for the 20nm-thick Au layer when the incident field is tuned at 800nm. The transmission curves are identical to those in Fig.3.

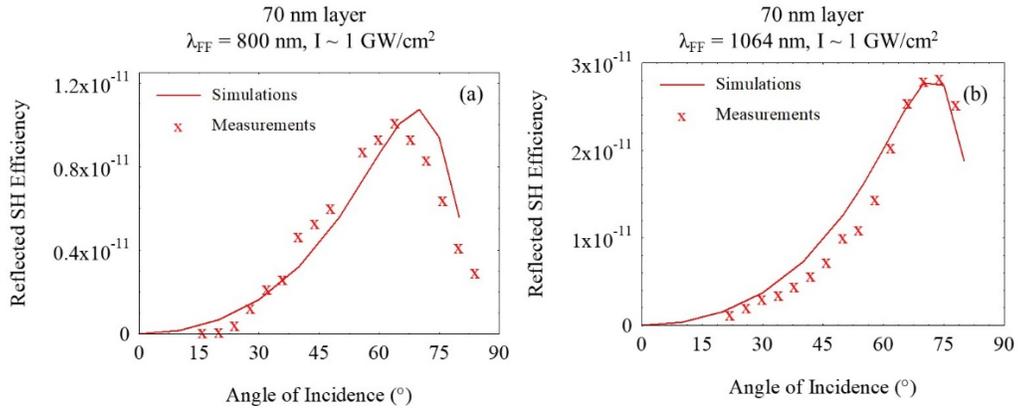

Fig. 5. Measured (red x symbols) and simulated (solid red curves) of reflected SHG efficiency as a function incident angle for the 70nm-thick Au layer when the carrier wavelength is (a) 800nm, and (b) 1064nm. We slightly adjust incident peak power density, effective electron mass and density and reproduce quite well the amplitudes and approximate peak locations in both cases.

As outlined above, a system described by combined Drude and Lorentz electrons illuminated by a pulsed pump laser source has at least two main THG sources: interband transitions (Lorentz resonances, with nonlinear interactions governed by the parameter $\tilde{\beta}$) and hot electrons, whose relative strength is determined by $\tilde{\Lambda}$. Other types of nonlinearities, such as a thermal response which may be slower, may be controlled and excluded simply by reducing pulse repetition rate, which we have done in our experiment so as to limit the kinds of nonlinearities active in our system. Any additional nonlinearity triggered on a time scale slower compared to the relatively fast electronic responses that we consider may be easily included by introducing appropriate terms in either or both Eqs.(1-2), depending on the origin of the process. Therefore, as written, Eqs.(1-2) fully account for the mixing of linear and nonlinear material dispersions due to the combined action of free and bound electrons, with modifications induced by the presence of nonlocal and magnetic effects. We may then use these equations, the macroscopic Maxwell's equations, and the constitutive relations to extract the nonlinear dispersive properties of the system. The approach is new but straightforward, and will be outlined in details in a separate effort. Presently we limit ourselves to presenting the results.

In Fig.6 we plot the predicted, reflected THG spectral response for 100fs pulses incident at 55°. The transmitted signal is predicted to display similar behavior. In this example the parameters are chosen arbitrarily so that each nonlinear contribution yields similar THG conversion efficiency when the other is turned off. Then we combine them together to ascertain the interplay between these two different TH sources. The figure suggests that separately, each type of third order nonlinearity yields qualitatively and quantitatively a similar response, with a TH peak for pump (TH) wavelength of ~600nm (200nm). However, their combined response redshifts the TH peak. This prediction runs counter to intuition because an increased free electron density should blueshift the plasma frequency, with expectedly similar outcome for the TH peak. It is obvious that the two components interfere and conspire to instead redshift the peak, an effect that encapsulates a cautionary tale for any experimental result that may be cavalierly extrapolated without the benefit of proper assumptions and theoretical support.

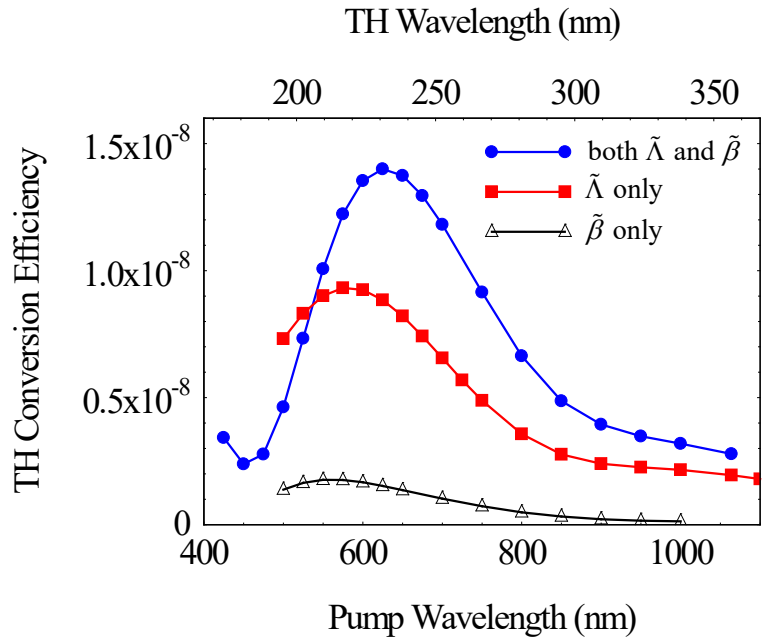

Fig. 6. Prediction of reflected THG conversion efficiency vs. wavelength for pump pulses incident at 55°.

The transmitted angular dependence of THG was recorded from the 20nm-thick gold layer using the 1064nm laser source. The repetition rate was varied to insure thermal effects were not excited. Measured and simulated THG power conversion efficiencies are shown in Fig.7.

The blue * markers denote the transmitted, TM-polarized TH for a TM-polarized incident field. The red x markers represent the TE-polarized TH signal for a TE-polarized pump. The angular dependence of THG is different compared to SHG because the inherent nonlinearity in this case is triggered by the bulk, with conversion efficiencies of order of $10^{-8}$. A scaled coefficient $\tilde{\beta} = 10^{-8}$ alone yields conversion efficiencies of order $10^{-11}$, clearly inadequate to explain our observations. Instead, what is required to reproduce the conversion efficiencies that we observe is the introduction of $\tilde{\Lambda} \approx 3 \times 10^{-8}$. Oscillatory behavior at small angles is likely due to the sensitivity of the set-up [25]. Nevertheless, the figure once again suggests that both efficiency and angular response are well-predicted by our model. The question that as yet remains unresolved is about the magnitude of the effective third order dispersive properties of the system.

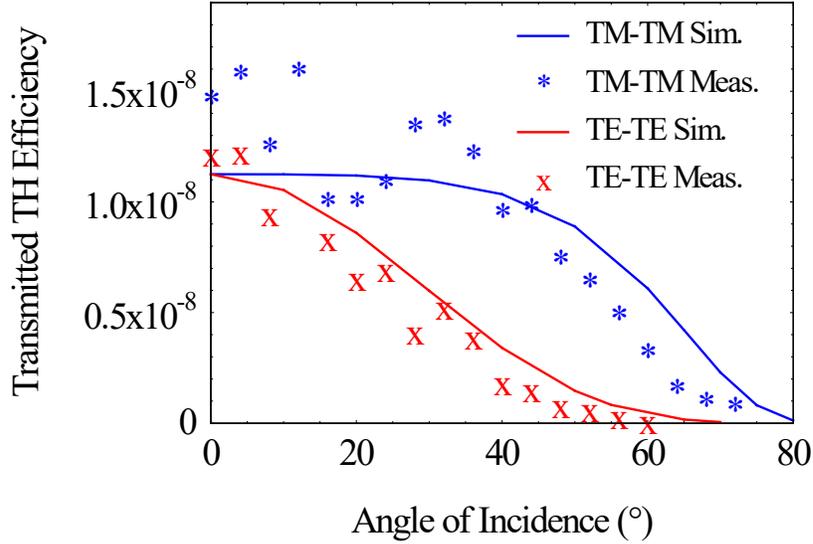

Fig. 7. Measured and simulated transmitted THG efficiency as a function of incident angle for TM- and TE-polarized incident pump pulses tuned to 1064nm, incident on a 20nm-thick gold layer. In the legend, TM-TM indicates the polarization of pump and generated TH, respectively, and similarly for TE-TE.

In Fig.8 we display the calculated nonlinear pump dispersions of a 20nm-thick gold layer. The solid curves represent the analytical solution of $\chi_\omega^{(3)}$, derived in a manner identical to that described for a single nonlinear oscillator [21, 26], assuming undepleted pump in the local approximation. The $\chi_\omega^{(3)}$ thus includes the superposition of two bound electron contributions that peak in the UV portion of the spectrum, and a free electron component that mirrors the dispersion of gold, with a real part that becomes increasingly negative at long wavelengths. This behavior should not be misconstrued as a trend at long wavelengths. Indeed, mirror-like behavior leads to reduced penetration depths and fields that are increasingly confined near the surface of the metal. Care should be exercised when performing and interpreting the results of the calculations, which include averaging the local fields over the thickness of the given layer. With this in mind, the markers-only curves stand for the total $\chi_\omega^{(3)}$ calculated at normal incidence using 100fs pulses (or longer), extracted using the macroscopic constitutive relations. They are in remarkable agreement with the analytical solutions, suggesting in this case minimal impact from nonlocal effects. The numerical method that reproduces and confirms the analytical results is illustrated in reference [23] specifically for the retrieval of the *linear* optical properties of an indium tin oxide nanolayer subject to nonlocal effects. Extension to the reconstruction of the nonlinear dispersion curves is straightforward but the details will be left for a future effort. Suffice it to say here that, using the constitutive relations, the dielectric constant may be expressed in terms of the instantaneous values of the fields averaged over the entire layer [23], as follows: $\varepsilon \approx 1 + 4\pi P/E$. Then the effective $\chi_\omega^{(3)}$ may be extracted by performing two separate calculations, in the linear and nonlinear regimes, and by taking the difference.

In the 1μm range, the *bound electron only* contribution to $\chi_\omega^{(3)}$, which is triggered by $\tilde{\beta}$, is positive and of order $10^{-20}\,(\text{m/V})^2$. Adding the hot electron component boosts the third order nonlinear coefficient in excess of two orders of magnitude to a predicted value of approximately $\chi_\omega^{(3)}(1064\text{nm}) \approx -3.7\times 10^{-18} + i\, 2.1\times 10^{-19}\,(\text{m/V})^2$. This value is estimated without ambiguity based exclusively on the experimentally observed THG conversion efficiency. This conclusion is straightforward because the parameters $\tilde{\beta}$ and $\tilde{\Lambda}$ are the only coefficients present in the equations of motion (not $\chi_\omega^{(3)}$ or $\chi_{3\omega}^{(3)}$), each is dominant within a given wavelength range, both preserve their respective nonlinear dispersions, and indeed determine the dynamics of third order nonlinear processes in their entirety: in the undepleted pump approximation, self-phase modulation and nonlinear absorption of the pump are proportional to $\tilde{\beta}|\mathbf{P}_\omega|^2\mathbf{P}_\omega$ and $\tilde{\Lambda}|\mathbf{E}_\omega|^2\mathbf{E}_\omega$, THG is proportional to $\tilde{\beta}(\mathbf{P}_\omega\cdot\mathbf{P}_\omega)\mathbf{P}_\omega$ and $\tilde{\Lambda}(\mathbf{E}_\omega\cdot\mathbf{E}_\omega)\mathbf{E}_\omega$, where the fields represent envelope functions. Standard z-scan measurements used to retrieve the nonlinear susceptibility in an absorptive medium are more complicated, require precise measurement in both open and closed aperture configurations, and slower, thermal effects may come into play [19]. As shown in reference [15], the common approximation where $\chi_\omega^{(3)}$ is related directly to the nonlinear refractive index $n_2$ and to the two-photon absorption coefficients $\alpha_2$, respectively, is invalid [27]. In contrast, by preserving linear and nonlinear dispersions, our model allows us to directly retrieve the nonlinear susceptibility without resorting to any type of transformation, and provides physical insight into the roles different sources may play, i.e. bound and hot electron contributions to the nonlinear susceptibility. We note that the extraction of $\chi_{3\omega}^{(3)}$ from the model is also easily accomplished, but not as straightforward, and the issue will be taken up at a later time. However, suffice it to say here that the result, shown in Fig.9, demonstrates that once again the analytical solutions and our calculations are in excellent agreement, suggesting that $\chi_{3\omega}^{(3)}(355\text{nm}) \approx -1.5\times10^{-19} + i1.1\times10^{-20}\,(\text{m/V})^2$. Derivations and methods used to depict the nonlinear susceptibilities shown in Figs. 8 and 9 are beyond the scope of this work, and will be presented elsewhere.

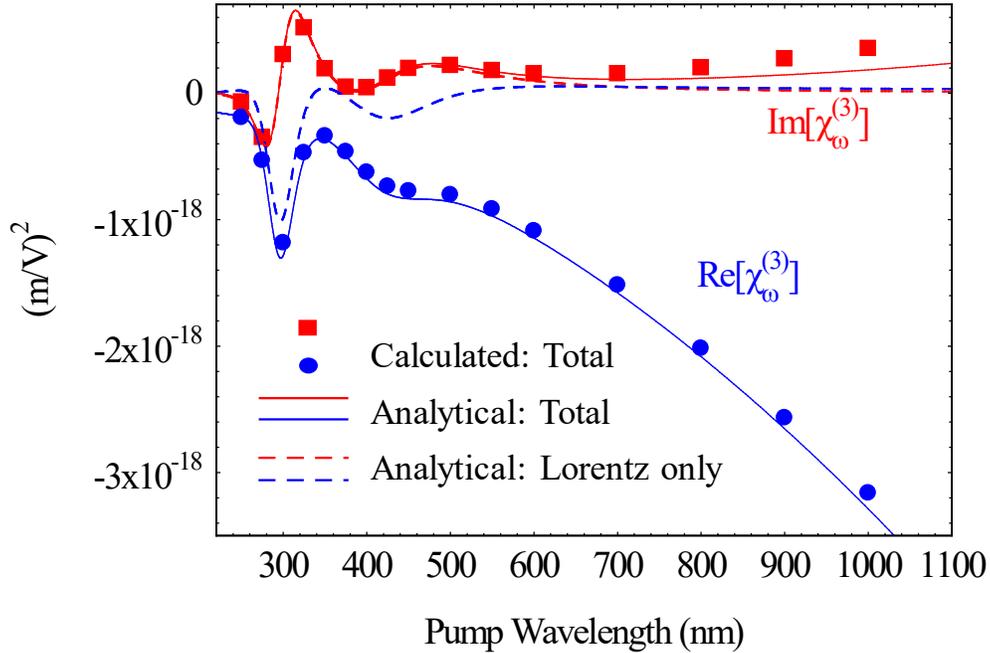

Fig. 8. Analytical and calculated nonlinear dispersion for the combined contributions of one Drude and two Lorentz components to the nonlinear third order susceptibility of gold, illuminated with low-repetition rate, picosecond pulses. At 1064nm, the increased free electron density augments the effective nonlinear coefficient by more than two orders of magnitude compared to the bound electron contributions only, making the real part of the susceptibility of order $-10^{-18}$ to $-10^{-17}\,(\text{m/V})^2$.

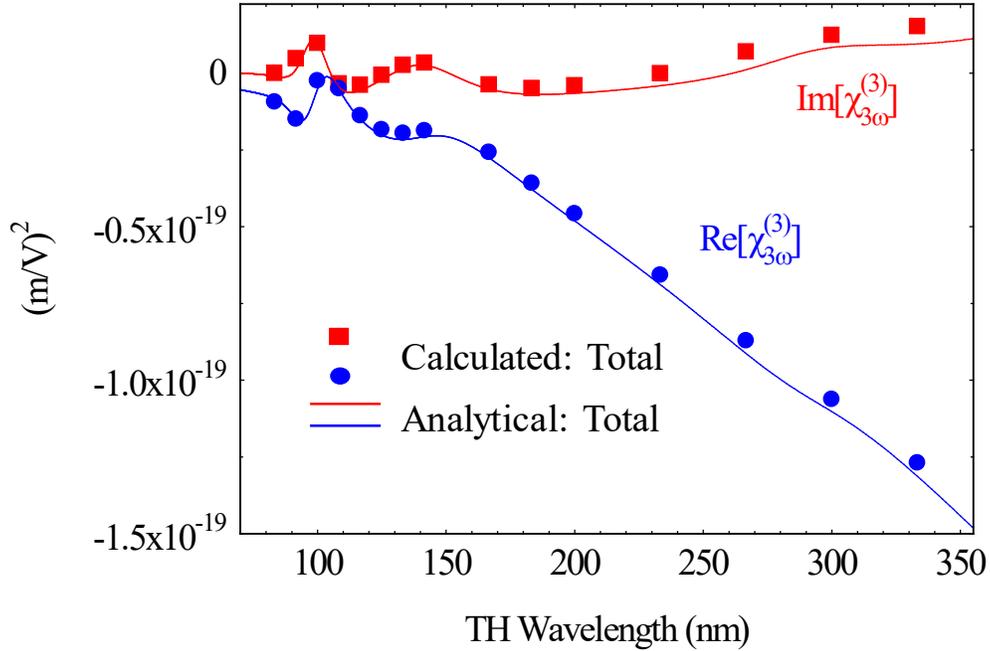

Fig. 9. Analytical and calculated total nonlinear dispersion $\chi_{3\omega}^{(3)}$ for the combined contributions of one Drude and two Lorentz components to the nonlinear third order susceptibility of gold, illuminated with low-repetition rate, picosecond pulses.

## 4. Conclusions

In conclusion, we have reported experimental observations of SHG and THG in transparent and opaque gold nanolayers. A transparent, 20nm-thick layer is used to monitor transmitted and reflected SH signals, as the fields carry information about combined surface a volume currents excited on and inside the sample. On the other hand, the opaque, 70nm-thick gold layer supports mostly surface currents, with reflected conversion efficiencies that are largest at larger angles. We use a microscopic, hydrodynamic approach to model the light-matter interactions that makes no assumptions about effective surface or volume nonlinearities, and relies instead on temporal and spatial derivatives to determine the relative magnitudes of surface and volume contributions. In doing so, we find unprecedented, remarkable agreement with experimental observations. We also report observations of TM- and TE-polarized, transmitted THG from the 20nm-thick gold layer. We attribute the generated TH signal mostly to hot electron dynamics, and use our model, together with the constitutive relations, to estimate $\chi^{(3)}_{\omega}(1064\text{nm}) \approx (-3.7 + i\,0.21) \times 10^{-18}\,(\text{m/V})^2$ and $\chi^{(3)}_{3\omega}(355\text{nm}) \approx (-2.43 + i0.13) \times 10^{-19}\,(\text{m/V})^2$, based solely on experimental values of THG power conversion efficiency. The results thus suggest one may retrieve the value of the overall third order dispersion functions without the need to implement a z-scan set-up, provided nonlinear dispersion is properly taken into account in the model. Finally, we note that intensity-driven increases in free electron density do not signifcantly alter the efficiencies of second order processes.

## Acknowledgment

LRS, JT and CC acknowledge financial support from RDECOM Grant W911NF-18-1-0126 from the International Technology Center-Atlantic, and from the Spanish Ministerio de Ciencia, Innovación y Universidades (project no. PID2019-105089GB-I00). MAV and DdC acknowledge financial support from Army Research Laboratory Cooperative Agreement Number W911NF-20-2-0078 from the International Technology Center-Atlantic. N.A. and M.S. acknowledge discussions with D. D. Smith.